\documentclass[journal=jpcl,manuscript=article]{achemso}

\usepackage[version=3]{mhchem} 
\usepackage{amssymb}
\usepackage[normalem]{ulem}
\usepackage[dvipsnames]{xcolor}

\newcommand{\eb}[1]{{\color{black}{#1}}}

\newcommand{\onlinecite}[1]{\hspace{-1 ex} \nocite{#1}\citenum{#1}}
\DeclareUnicodeCharacter{2212}{-}

\author{Eric R. Beyerle}
\affiliation[University of Maryland, College Park]
{Institute for Physical Science and Technology, University of Maryland, College Park, MD}
\email{ebeyerle@umd.edu}
\author{Pratyush Tiwary}
\affiliation[University of Maryland, College Park]
{Institute for Physical Science and Technology, University of Maryland, College Park, MD}
\alsoaffiliation[University of Maryland, College Park]
{Department of Chemistry, University of Maryland, College Park, MD}

\title[Thermodynamically Optimized Machine-learned Reaction Coordinates for Hydrophobic Ligand Dissociation]
  {Thermodynamically Optimized Machine-learned Reaction Coordinates for Hydrophobic Ligand Dissociation}

\abbreviations{SPIB, MD, CVs}
\keywords{American Chemical Society, \LaTeX}

\begin{document}

\begin{tocentry}

\includegraphics[width=3.25in, height=1.75in]{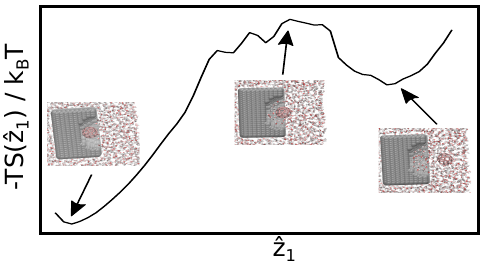}

\end{tocentry}

\begin{abstract}

Ligand unbinding is mediated by the free energy change, which has intertwined contributions from both energy and entropy. It is important but not easy to quantify their individual contributions. We model hydrophobic ligand unbinding for two systems, a methane particle and a C$_{60}$ fullerene, both unbinding from hydrophobic pockets in all-atom water. By using a modified deep learning framework, we learn a thermodynamically optimized reaction coordinate to describe hydrophobic ligand dissociation for both systems. Interpretation of these reaction coordinates reveals the roles of entropic and enthalpic forces as ligand and pocket sizes change. Irrespective of the contrasting roles of energy and entropy, we also find that for both the systems the transition from the bound to unbound states is driven primarily by solvation of the pocket and ligand, independent of ligand size. Our framework thus gives useful thermodynamic insight into hydrophobic ligand dissociation problems that are otherwise difficult to glean.
  
\end{abstract}

Ligand dissociation is an important process driving conformational change and functionality of proteins and other macromolecules\cite{Alberts2007}. One of the most important examples of ligand unbinding is the dissociation of inhibitory drug from a target protein molecule \cite{Badaoui2022}. Experiments are excellent for determining the thermodynamics \cite{Akke2012} an kinetics of ligand unbinding\cite{Amaral2017}, but they can lack direct mechanistic details of the unbinding event at the atomic level. As such, a common approach to gain information regarding the dissociation mechanism at the atomic spatial scale and with high temporal resolution is through the use of atomistic molecular dynamics (MD) simulations\cite{Tiwary2017}.  However, for ligands with a small dissociation constant, the residences times in the receptor are prohibitively long for study with atomistic MD, and, to study the dissociation process computationally, enhanced sampling procedures are required \cite{Badaoui2022,Ansari2022,Ccoa2022,Limongelli2013}.

Generally, once an MD simulation has achieved adequate sampling of the association-dissociation process, the simulation trajectory statistics is used to model the effective reaction coordinate (RC) for describing the dissociation event. While the simplest reaction coordinate one generally considers is the distance of the ligand from the binding site, it is not the most informative RC. This is because it accounts for only a single degree of freedom and ignores contributions from the solvent and any relevant internal degrees of freedom the system may possess. Variational methods can be utilized to optimize the RC to elucidate the details of the dissociation mechanism and kinetics \cite{Tiwary2017}. Furthermore, these more detailed RCs will be more informative regarding how the ligand and binding site behave at the transition state straddling the bound and unbound states.

Here, we model the ligand-receptor dissociation process by utilizing two model systems of hydrophobic binding: a united atom methane particle \cite{Setny2007, Baron2012,Baron2010,Setny2010} and a C${60}$ Buckminster fullerene \cite{Ccoa2022,Tiwary2015,Ahalawat2020} binding to a hydrophobic cavity that interacts with the ligand and surrounding TIP4P water solvent via dispersion interactions only; visual representations of these systems are given in in Figure S1.  We choose to study two systems of hydrophobic dissociation because it is known \cite{Lum1999,Rego2022,Weiss2017} that both the size and shape of ligand and cavity affect the thermodynamics, kinetics, and mechanism of unbinding. There are many methods available to develop potential RCs for studying this process, including using maximum likelihood of sampling paths \cite{Peters2006}, estimating transfer matrices \cite{Bowman2013}, principal\cite{Jolliffe2002} and independent\cite{Hyvarinen2001} component analyses, and machine learning based approaches \cite{Bittracher2023,Bonati2021,Chen2018,Wehmeyer2018,Hernandez2018,Varolgunecs2020}, among others. Here we find RCs for describing the unbinding process in both systems using the state predictive information bottleneck (SPIB) method\cite{Wang2021a,Mehdi2022,Beyerle2022}, a deep learning based method that finds non-linear RCs using a variational autoencoder (VAE)\cite{Kingma2014} architecture. We choose this framework due to its ability to accurately predict the metastable states of a system \cite{Wang2021a,Mehdi2022}, model the committor function \cite{Wang2021a,Beyerle2022}, and learn the effective driving mechanisms for rare events in solution\cite{Zou2023,Mehdi2022}. Furthermore, this framework allows for introducing thermodynamic intuition into machine learning of the reaction coordinates. As we show in this letter, this extension is a powerful way to quantify and optimize the enthalpic and entropic contributions to the thermodynamic barriers from simulations at just a single temperature.

The thermodynamics of unbinding for a similar methane system studied here has been previously elucidated in great detail previously \cite{Baron2012,Baron2010,Setny2010}, including a separation of the contributions of the free energy of unbinding into its energetic and entropy contributions. Given the sensitivity of the thermodynamics of hydrophobic association and dissociation to hydrophobe shape and size\cite{Rego2022}, we modify the SPIB loss function with an extra term in the spirit of the EncoderMap approach \cite{Lemke2019} to encourage the separation of the energy and entropy barriers surmounted during the dissociation process into two separate reaction coordinates. This modification of the SPIB loss function allows the architecture to effectively learn the relevant thermodynamic profiles of hydrophobic unbinding, explicitly adding physics into the neural network's learning procedure.

For both systems, we find that the free-energy barrier to dissociation is overall dominated by an entropy barrier. However for methane, there is also a small energy barrier impeding dissociation while fullerene dissociation is entirely downhill in energy. Furthermore, for both these systems, we find that a one-dimensional RC space is adequate for capturing the thermodynamics of unbinding. This result is due to the dominance of the entropy barrier to the unbinding free-energy barrier and a lack of a significant energy barrier to unbinding along the learned RC in both cases. Modifying the SPIB to explicitly account for energy and entropy barriers along the RC is critical for finding these thermodynamic barriers; without it, the SPIB returns a \eb{thermodynamically ignorant RC that my miss critical intricacies regarding the ligand unbinding mechanism.}

\begin{figure}[htb] 
\center
\includegraphics[width=0.9\linewidth]{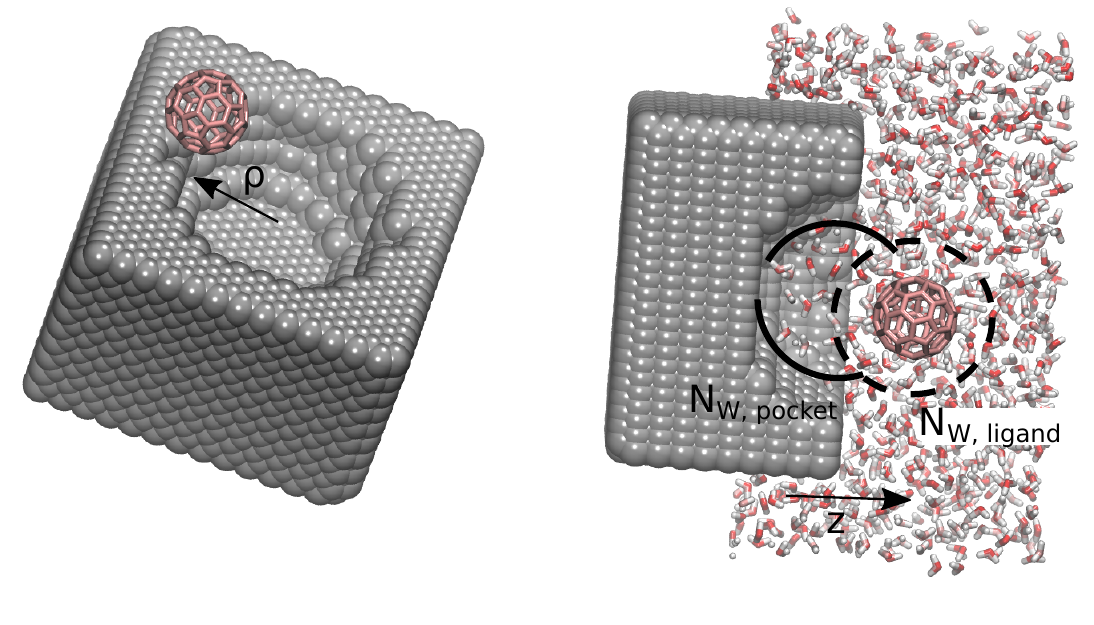}
\caption{Schematic definitions of the input features to the SPIB analysis for RC discovery in the fullerene system; the definitions are analogous for the methane system. The schematic on the left illustrates the definition of the radial distance from the center of the pocket, $\rho=\sqrt{x^2 + y^2}$, with $x$ and $y$ both defined with respect to the center-of-mass of the atoms lining the pocket. The schematic on the right shows the definition of both the pocket- and ligand-water coordination numbers, N$_{\text{W, pocket}}$ and N$_{W, ligand}$, respectively, as well as the z-distance from the pocket, z. Both images are created using VMD v1.9.3\cite{Humphrey1996}.}
\label{fig:OPs}
\end{figure}

This study is not the first to study non-trivial RCs for the fullerene system. In Ref. \onlinecite{Ahalawat2020}, the authors utilize a combination of time-lagged independent component analysis (tICA)\cite{Schwantes2013,Perez-Hernandez2013} and Markov state models (MSMs)\cite{Klus2018} to discover novel RCs for fullerene dissociation from an identical hydrophobic pocket as presented here. An optimized, in the sense of maximal spectral gap separation, one-dimensional RC was developed using the spectral gap optimization of order parameters (SGOOP)\cite{Tiwary2016wet} and reweighted autoencoder variational Bayes (RAVE)\cite{Ribeiro2018}, the precursor to SPIB, techniques. Both methods found that the ligand's z-distance from the pocket, with the hydration state of the pocket contributing little to the optimized RC. However, in both cases, only three input features were considered while constructing the RC: the z-distance of the ligand from the pocket; the radial distance of the ligand from the center of the pocket, $\rho$; and the hydration state of the pocket, N$_{\text{W, pocket}}$. For the methane-pocket system, RCs different from the ligand's z-distance from the pocket have not been examined in detail. Here, we use a richer six input feature basis, described in Figure \ref{fig:OPs} for the fullerene system, to learn a non-linear RC that optimizes the entropy barrier using the augmented SPIB formalism. Explicitly, the six features input to the SPIB are the $(x, y, z)$ coordinates of the ligand in the reference frame of the pocket; radial distance of the ligand from the center of the pocket, $\rho=\sqrt{x^2 + y^2}$; and the solvation state of the pocket, N$_{\text{W, pocket}}$, and the ligand, either methane, N$_{\text{W, methane}}$, or the fullerene,  N$_{\text{W, fullerene}}$. Since both systems are constrained along the $\rho$ coordinate, the x, y, and $\rho$ input features serve essentially as noisy features.

The SPIB architecture is based on the variational autoencoder (VAE) architecture\cite{Kingma2014}, with the addition of a lagtime $\tau$ to the loss function and the prediction of state labels in place of reconstruction of the original input data. These augmentations to the original VAE give a model that minimizes the following loss function, which is similar in structure to the variational information bottleneck\cite{Alemi2017}:
\begin{align}
\mathcal{L}=&\mathbb{E}_{p_{\theta}(\mathbf{z}|\mathbf{x})}\left[-\log\left(q_{\theta}(\mathbf{y}(t + \tau)|\mathbf{z}(t))\right) + \beta\log\left(\frac{p_{\theta}(\mathbf{z}(t)|\mathbf{x}(t))}{r_{\theta}(\mathbf{z})}\right)\right]\notag\\
=&\mathbb{E}_{q_{\theta}(\mathbf{z}|\mathbf{x})}\left[-\log\left(p_{\theta}(\mathbf{y}(t + \tau)|\mathbf{z}(t))\right)\right]+\beta\text{D}_{\text{KL}}\left(p_{\theta}(\mathbf{z}(t)|\mathbf{x}(t))||r_{\theta}(\mathbf{z})\right).
    \label{eq:loss}
\end{align}
In eq. \ref{eq:loss}, $p_{\theta}(\mathbf{z}|\mathbf{x})$ is the encoder generating the latent space $\mathbf{z}$ from the input features $\mathbf{x}$, $q_{\theta}(\mathbf{y}(t + \tau)| \mathbf{z}(t))$ is the decoder generating the predicted metastable state, $\mathbf{y}$, at a lagtime $\tau$ given the observed value of the RC at time $t$, $\mathbf{z}(t)$, $r_{\theta}(\mathbf{z})$ is an assumed mulitmodal prior distribution of $\mathbf{z}$, and $\theta$ denotes the set of all learnable parameters of the model.
The optimal RC space $\mathbf{z}$ minimizing eq. \ref{eq:loss} is a set of RCs that optimally predicts the coarse-grained dynamics in the state space $\mathbf{y}$ at a lagtime $\tau$ in the future while simultaneously minimizing the Kullback-Leibler divergence from an initial, assumed prior distribution $r_{\theta}(\mathbf{z})$, with the trade-off given by the hyperparameter $\beta$. The prior for this model is taken as a set of encoded representations from each of the separate states in the state space $\mathbf{y}$ from the training data extracted from the biased MD simulations, which prevents posterior mode collapse\cite{Bond2021}. Further details regarding the SPIB theory and architecture can be found in previous publications\cite{Wang2021a,Mehdi2022,Beyerle2022} and the associated GitHub repository (https://github.com/tiwarylab/State-Predictive-Information-Bottleneck). Specific details regarding the SPIB training for the two model systems presented here are listed in Table \ref{tab:spib}; further details regarding the training are given in the SI.
\begin{table}[h!]
{
   \centering
   \caption{SPIB Parameters for the Methane and fullerene Systems}
   \label{tab:spib}

   \begin{tabular}[c]{l c c c }
   System & $\tau$, ps & $\beta$ & lr$^a$\\ \hline
   Methane & 5 & 1$\times$10$^{-2}$ & 0.0001 \\
   fullerene & 10 & 1$\times$10$^{-2}$ & 0.0001 \\ 
   \multicolumn{4}{l}{\textsuperscript{\emph{a}}{\scriptsize Learning Rate}}\\
   \end{tabular}
   }
\end{table}

To encourage the RC space discovered by SPIB to have one RC that maximizes the entropy barrier along and another that maximizes the energy barrier, we introduce an extra physics-based regularization term, $\gamma f(\mathbf{z})$, to the SPIB loss function to generate the following:

\begin{align}
\mathcal{L}=&\mathbb{E}_{p_{\theta}(\mathbf{z}|\mathbf{x})}\left[-\log\left(q_{\theta}(\mathbf{y}(t + \tau)|\mathbf{z}(t))\right)\right]+\beta\text{D}_{\text{KL}}\left(p_{\theta}(\mathbf{z}(t)|\mathbf{x}(t))||r_{\theta}(\mathbf{z})\right) \notag\\
&- \gamma f(\mathbf{z}), 
    \label{eq:loss2}
\end{align}
with $\gamma$ an extra hyperparameter akin to $\beta$ and $f(\mathbf{z})$ an arbitrary function of the RC space. To optimize the entropy and energy profiles in the learned RC space, $f(\mathbf{z})$ should be set in the following form:
\[f(\mathbf{z})=\max_{\text{int }z_1}\left(-T\Delta \text{S}(z_1)\right) + \max_{\text{int }z_2}\left(\Delta \text{U}(z_2)\right) \ \ \ (\text{a}),\]
where $z_1$ and $z_2$ are the two components of $\mathbf{z}$, $-T\Delta \text{S}(z_1)$ the entropy barrier along $z_1$, and $\Delta\text{U}(z_2)$ the energy barrier along $z_2$. The notation $\max_{\text{int } z_i}$ denotes that the barrier maximum is taken on the interior of the coordinate dimension spanned by $z_i$. That is, the boundaries of $z_i$ are ignored in the optimization of the thermodynamics input to the loss function eq. \ref{eq:loss2} to avoid spurious optimization of noisy barriers caused by poor sampling on the boundaries of $z_i$.

For the two hydrophobic systems studied here, we find that optimization of the energy barrier along $z_2$ produces a redundant coordinate with little extra information. From this we conclude that the role of energetic barriers in the dissociation process is dwarfed by the entropy contribution, which has been observed previously for the methane unbinding system \cite{Baron2010,Baron2012}. That is, in the two cases of hydrophobic ligand dissociation studied here, a single RC captures the significant free-energy barrier to unbinding, which is dominated by the entropy contribution, and the second RC optimized along the energy contribution can be ignored safely. As such, for the results shown in this Letter, we only learn a one-dimensional RC space subject to the added constraint
\[f(\mathbf{z})=f(z_1)=\max_{\text{int }z_1}\left(-T\Delta \text{S}(z_1)\right) \ \ \ (\text{b})\]
to simplify and robustify the RC learning process. In this case, the only advantage to thermodynamic training along the second RC is to serve as a regularization term improving the original SPIB's ability to optimize entropy and entropy barriers in the RC space.

The energy and entropy profiles along the RC are calculated using the geometric free energy\cite{Lelievre2010,Hartmann2011}, as described previously \cite{Beyerle2022}. Briefly, the free energy, energy, and entropy are calculated in one-shot from simulations run at a single temperature $T$ using following definitions:
\begin{equation}
G({\mathbf{z}})=-k_{B} T \ln \left(\int_{R^{n}} e^{ -U(\mathbf{x}) / k_{B} T}\right. \\
\left.\times \delta(\Phi(\mathbf{x})-{\mathbf{z}}) \operatorname{det}(\tilde{G})^{\frac{1}{2}} d \mathbf{x}\right)
    \label{eq:Gz}
\end{equation}

\begin{align}
\langle \text{U}({\mathbf{z}})\rangle_{\Sigma({\mathbf{z}})} := \text{U}({\mathbf{z}}) = \frac{1}{N_{\mathbf{z}}}\int &\text{U}(\mathbf{x})e^{-U(\mathbf{x})/k_BT} \notag \\ 
&\times\delta\left(\Phi(\mathbf{x}) - {\mathbf{z}}\right)\det\left(\tilde{G}\right)^{\frac{1}{2}}d\mathbf{x} 
\label{eq:Uz}
\end{align}

\begin{equation}
    -T\Delta\text{S}(\mathbf{z})=\Delta \text{G}(\mathbf{z}) - \Delta \text{U}(\mathbf{z}).
\end{equation}
In the above, $N_{\mathbf{z}}$ is a normalization constant:
\[N_{\mathbf{z}}=\sum_{k=1}^{N} I_{\Phi(x(k))} \operatorname{det}(\tilde{G})^{\frac{1}{2}},\]
$\tilde{G}$ is the gram matrix of the coordinate transformation induced by the SPIB encoder $\Phi(\mathbf{x})$, $\int_{\mathbb{R}^n}d\mathbf{x}\left(\cdots\right)$ denotes integration over the n-dimensional input space to the SPIB neural network, $\delta\left( \cdots\right)$ is the Dirac delta function, $\langle f(\mathbf{x}) \rangle_{\Sigma(\mathbf{z})}=\frac{1}{N_{\mathbf{z}}}\int d\mathbf{x} f(\mathbf{x})e^{-U(\mathbf{x})/k_BT}$ denotes the averaging of some generic function of the inputs $f(\mathbf{x})$ over a given level set $\Sigma(\mathbf{z})$ of $\mathbf{z}$. Finally, $I_{\Phi(x(k))}$ is an indicator function over $\Sigma({\mathbf{z}})$ which is equal to 1 if $\Phi(x(k))$ maps to $\Sigma({\mathbf{z}})$ and is equal to 0 otherwise. Finally, it should be noted that since we use the already reduced input feature space $\mathbf{x} \in \mathbb{R}^n$, with $n<N$, $N$ being the size of the system configuration space, the discovered RCs are $\emph{not}$ gauge invariant, in general.

For this SPIB variant to learn thermodynamic-barrier optimized RCs, it requires an input time series of the described input features with a constant timestep between them. Practically, the only way to generate such a time series is through the use of atomistic molecular dynamics (MD) trajectories. All MD simulations reported here are performed using GROMACS 2021.4 \cite{Abraham2015} patched with PLUMED 2.8.0 \cite{Tribello2014}. The paraffin-like walls for the simulation with methane as the ligand are built using LAMMPS (build 10 March 2021) \cite{Plimpton2022}. For the fullerene system, the starting structure and GROMACS run files (topologies and PLUMED files) are taken from the GitHub repository corresponding to Ref. \onlinecite{Ccoa2022} (https://github.com/hocky-research-group/PenaUnbindingPaper). Both systems are solvated in TIP4P water, and Lennard-Jones interactions between particles are evaluated using Lorentz-Berthelot mixing rules. These interaction parameters are given in Tables \ref{tab:methane} and \ref{tab:buckyball}. More explicit details regarding the MD simulations for both the methane and fullerene simulations are given in the SI.

Since ligand unbinding is typically a rare event on the molecular scale, we utilize well-tempered metadynamics (WTMetaD)\cite{Barducci2008} to accelerate ligand unbinding from the pocket. Both the methane and fullerene simulations are biased along the z coordinate alone with harmonic position restraints with a spring constant equal to 418.4 kJ /(mol nm nm) in the $\rho$ coordinate. The biased production runs total 76.1 ns and 94.2 ns in length for the methane and fullerene simulations, respectively. These simulation timescale has previously been shown to be adequate for converging the free energy for both systems using either umbrella sampling \cite{Baron2012,Setny2007} or WTMetaD \cite{Ccoa2022,Tiwary2016wet,Ribeiro2018,Tiwary2015}. For WTMetaD, the width of the deposited Gaussians is found by running a short, unbiased simulation and calculating the standard deviation of the z coordinate in each simulation. Further details regarding the WTMetaD parameters are given in the SI, and the GROMACS input files and PLUMED files required for reproducing the biased simulations for both the methane and fullerne systems are available on GitHub (https://github.com/tiwarylab/hydrophobic-ligand-dissociation) and the PLUMED nest (https://plumed.org/nest/eggs/XXX).
\begin{table}[h!]
{
   \centering
   \caption{Lennard-Jones parameterization for the methane binding system}
   \label{tab:methane}

   \begin{tabular}[c]{l c c}
   System & $\epsilon$, kJ/mol & $\sigma$, nm\\ \hline

   Methane & 1.2301 & 0.373 \\ 
   Wall & 0.0024 & 0.4152 \\
   Pocket & 0.008 & 0.4152 \\
   TIP4P$^a$ & 0.6485 & 0.3154 \\
   \multicolumn{3}{l}{\textsuperscript{\emph{a}}{\scriptsize Interaction site on the oxygen atom}}\\
   \end{tabular}
   }
\end{table}
\begin{table}[h!]
{
   \centering
   \caption{Lennard-Jones parameterization for the fullerene binding system}
   \label{tab:buckyball}

   \begin{tabular}[c]{l c c}
   System & $\epsilon$, kJ/mol & $\sigma$, nm\\ \hline

   fullerene & 0.2761 & 0.35 \\ 
   Wall & 0.0024 & 0.4152 \\
   Pocket & 0.008 & 0.4152 \\
   TIP4P$^a$ & 0.6485 & 0.3154\\
   \multicolumn{3}{l}{\textsuperscript{\emph{a}}{\scriptsize Interaction site on the oxygen atom}}\\
   \end{tabular}
   }
\end{table}

\begin{figure}[tp] 
\center
\includegraphics[width=0.9\linewidth]{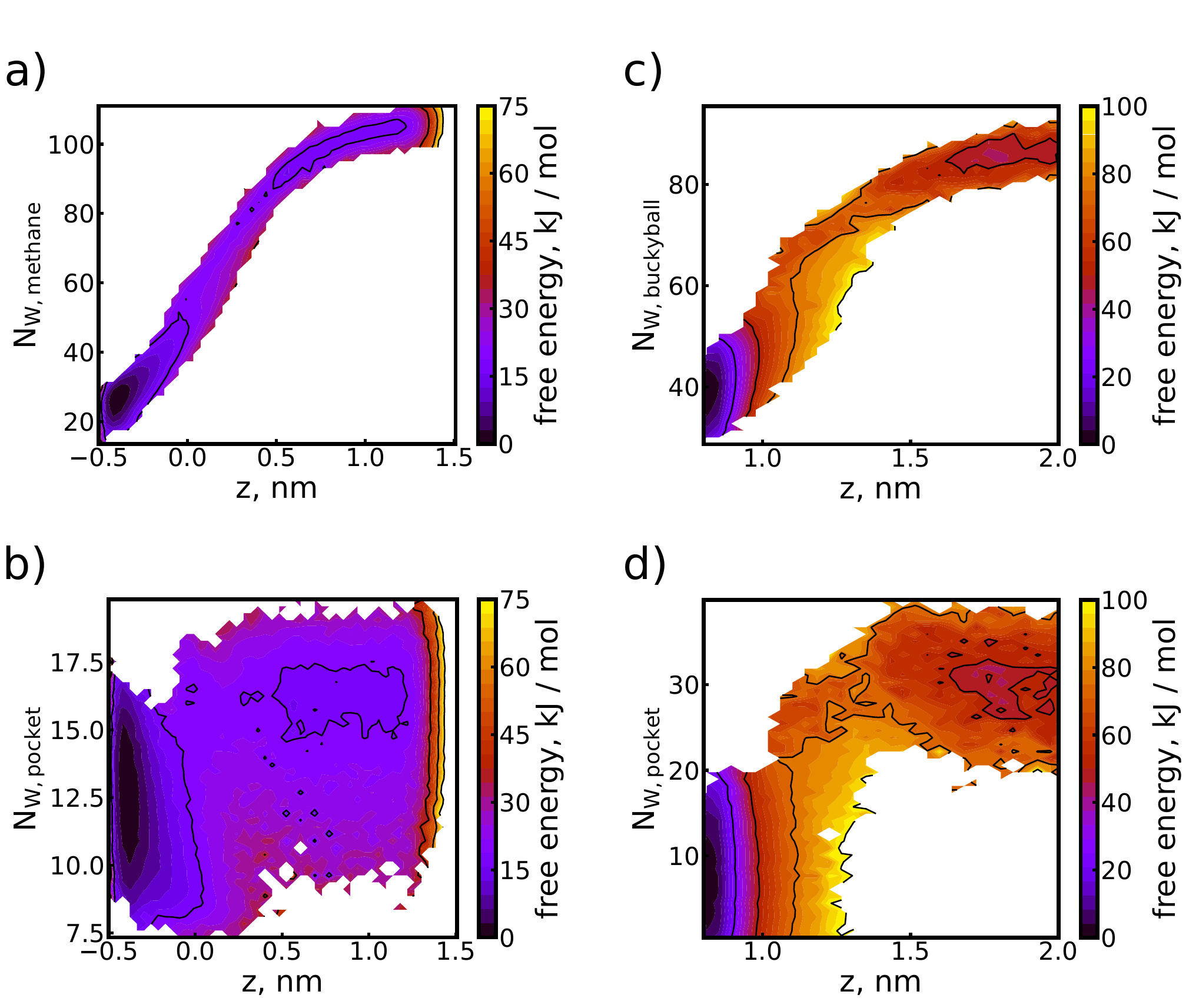}
\caption{Two-dimensional free-energy surfaces for methane as a function of a) (z, N$_{\text{W, methane}}$) and b) (z, N$_{\text{W, pocket}}$) and the fullerene as a function of c) (z, N$_{\text{W, fullerene}}$) and d) (z, N$_{\text{W, pocket}}$). }
\label{fig:fes}
\end{figure}

The WTMetaD trajectories from both systems show good sampling of both the ligand bound and the ligand unbound states; the sampling can be quantified by plotting free-energy surfaces in the space of the z-distance and hydration states of the pocket and ligand. Figure \ref{fig:fes} show two-dimensional free-energy surfaces for the methane (Figure \ref{fig:fes}a and \ref{fig:fes}b) and fullerene (Figure \ref{fig:fes}a and \ref{fig:fes}b) systems as functions of the z-distance from the pocket and a solvation coordinate, either the solvation state of the solute (Figure \ref{fig:fes}a and \ref{fig:fes}c) or the pocket (Figure \ref{fig:fes}b and \ref{fig:fes}d). While the plots for both solutes are qualitatively similar in that as the solute moves further from the pocket, on average, both its solvation state and that of the pocket increases, there are quantitative differences due to the differing geometries of the pocket and the solutes. For methane, due to its smaller size, its hydration is sigmoidal-shaped as a function of z while the hydration of the fullerene is more linear as a function of z.  Furthermore, for the methane system, the solvation of the pocket is a nontrivial function of z, (Figure S2 and Ref. \onlinecite{Baron2012}) while for the fullerene system, it is monotonically increasing with z until an insignificant maximum when the fullerene is in the bulk solvent (Figure S2), indicating different drying effects upon unbinding caused by the differing system geometries.

\begin{figure}[tp] 
\center
\includegraphics[width=0.9\linewidth]{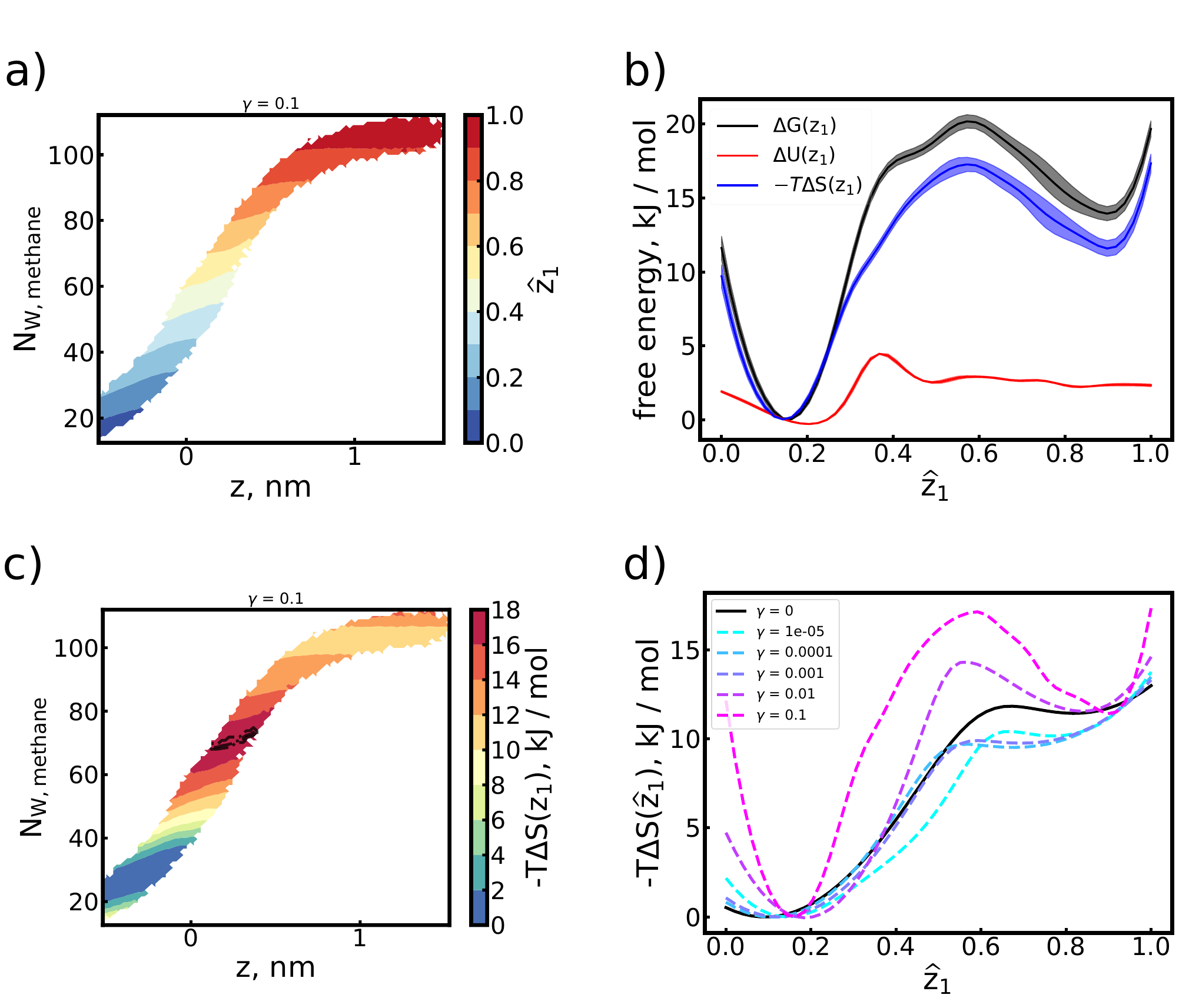}
\caption{a) Projection of the learned SPIB one-dimensional RC z$_1$ onto the space spanned by (z, N$_{\text{W, methane}}$) features. b) Thermodynamic profile along a normalized version of z$_1$, $\widehat{\text{z}}_1$, with the decomposition into energy, entropy, and free energy shown. Shaded error bars correspond to one standard error of the mean calculated by dividing the trajectory into four blocks. c) Projection of the entropy term -T$\Delta$S(z$_1$) to the free-energy profile along z$_1$ onto the space spanned by (z, N$_{\text{W, methane}}$). The dashed contour line shows the contour between the two metastable states discovered by the SPIB RC, which nearly coincides with the entropy barrier along the SPIB RC also shown in b). d) Evolution of the entropy contribution to the free-energy profile along the learned reaction coordinate z$_1$ as a function of the $\gamma$ hyperparameter controlling the relative contribution of the entropy penalty to the SPIB loss function. In this instance, the largest value of the hyperparameter $\gamma$ gives an RC with the largest entropy barrier along the profile.}
\label{fig:spib_methane}
\end{figure}

The above noted differences in the solvation behavior of the pocket and solute cause differences in the learned reaction coordinates describing the unbinding process. For methane, the results of the SPIB analysis for discovering the unbinding reaction coordinate are given in Figure \ref{fig:spib_methane}. A normalized version of the learned RC shown in Figure \ref{fig:spib_methane}a is a non-trivial function of both z and N$_{\text{W, methane}}$, showing that measuring the thermodynamics as a function of z alone may not be optimal. Figure \ref{fig:spib_methane}b shows that exiting the pocket along this learned RC requires the methane to first surmount a small energy barrier of around 5 kJ / mol. This energy barrier is followed by a much larger entropy barrier over 15 kJ / mol, which constitutes the majority of the free-energy penalty for unbinding. The peak of this significant entropy barrier coincides nicely with the border between the metastable states discovered by the SPIB method, as shown in Figure \ref{fig:spib_methane}c. Mechanistically, we find through direct observation of simulation frames on the border between the metastable states roughly corresponds to times when the methane particle is located at the mouth of the pocket, near z = 0.0 nm, which also corresponds closely to the maximally dry state of the pocket. Thus, this entropy barrier is likely due to hydrophobic vacuum formation within the pocket.  A few of these frames sampled from the peak of the entropy barrier are shown explicitly in Figure S4. Concurrently, the energy barrier is likely caused by the non-monotonic wetting of the pocket as the methane particle moves to larger values of z because, as the methane samples the maximally dry pocket state, it gains no favorable interactions with the pocket but loses some with the pocket-bound waters. Since both pocket and ligand are purely hydrophobic, the hydrophobic effect generates both the energy and entropy barriers for dissociation of methane. 

Finally, Figure \ref{fig:spib_methane}d shows that the above analysis could not have been obtained without the use of physics based regularization of the machine learning loss function. The development of this entropy barrier is demonstrated in Figure \ref{fig:spib_methane}d, where the -T$\Delta$S(z$_1$) profile is shown as a function of the hyperparameter $\gamma$ governing the weight of the entropy barrier along the RC to the SPIB loss function, as given in eq. \ref{eq:loss2}. In general, increasing the value of $\gamma$ causes an increase in the entropy barrier along the RC, which, given that we are learning a one-dimensional RC, is the expected behavior when increasing the weight of the f($\mathbf{z}$) term in the SPIB loss function. Without the inclusion of this extra regularization term in the SPIB loss function, we find only one metastable state and an incorrect physical picture of ligand dissociation (Figure S5).

\begin{figure}[tp] 
\center
\includegraphics[width=0.9\linewidth]{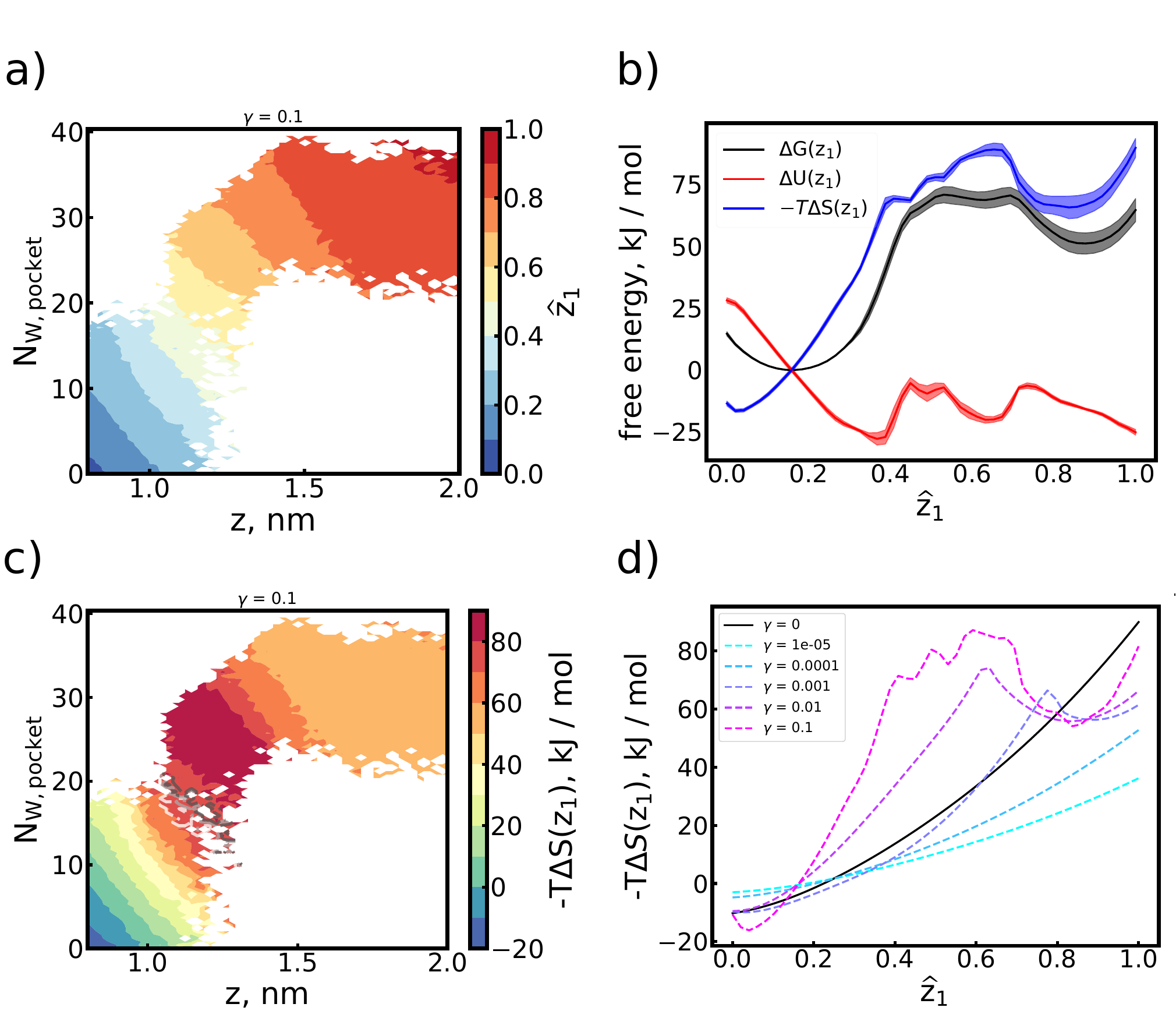}
\caption{a) Projection of the normalized learned one-dimensional SPIB RC onto the space spanned by (z, N$_{\text{W, pocket}}$) for the fullerene system. The RC takes high values when the fullerene is bound to the pocket and low values in the bulk solvent. b) Thermodynamic profile along the SPIB RC with the decomposition into energy, entropy, and free-energy shown. The vertical, dashed, grey line marks the approximate value of $\widehat{z}_1$ where the fullerene can be considered to be unbound and in the bulk solvent state. Shaded error bars correspond to one standard error of the mean calculated by dividing the trajectory into four blocks. c) Projection of the entropy term -T$\Delta$S(z$_1$) to the free-energy profile along z$_1$ onto the space spanned by (z, N$_{\text{W, pocket}}$). The dashed contour line shows the contours between the metastable states discovered by the SPIB RC, which nearly coincides with the onset of the entropy barrier along the SPIB RC in b). d) Evolution of the entropy contribution to the free-energy profile along the learned reaction coordinate z$_1$ as a function of the $\gamma$ hyperparameter controlling the relative contribution of the entropy penalty to the SPIB loss function. As is the case with methane unbinding, utilization of the $\gamma$ hyperparameter is required to discover an RC with a substantial entropy barrier.}
\label{fig:spib_fullerene}
\end{figure}

An analogous analysis for the SPIB-learned RC for fullerene unbinding when the $\gamma$ hyperparameter is set to 0.1 is given in Figure \ref{fig:spib_fullerene}. As with methane, the learned RC for fullerene unbinding, shown in Figure \ref{fig:spib_fullerene}a, reports on whether the ligand is bound or unbound and is a non-trivial function of z. Figure \ref{fig:spib_fullerene}b shows that the free-energy barrier to unbinding is due exclusively to entropy effects, with unbinding being downhill in energy along the RC, in contrast to the case for methane. The entropy -T$\Delta$S(z$_1$) is shown projected onto the (z, N$_{\text{W, pocket}}$) space in Figure \ref{fig:spib_fullerene}c, with the boundary between the bound and unbound states shown by the grey contour lines. This boundary lies near the peak of the entropy barrier, similar to the case of methane unbinding. This similarity is likely due to similarities in the physical origin of the entropy barrier, which corresponds to the de-wetting transition where waters start to fill the pocket (Figure S3) and occurs when the fullerene is poised at the mouth of the pocket (Figure S4).

Finally, Figure \ref{fig:spib_fullerene}d shows the entropy barrier along the learned RC as a function of the $\gamma$ hyperparameter. As $\gamma$ is increased beyond 10$^{-3}$, an entropy barrier appears and grows monotonically. In contrast to the case of methane unbinding, we do not require the $\gamma f(\mathbf{z})$ term in the SPIB loss function to discover multiple metastable states for fullerene unbinding (Figure S6), but increasing $\gamma$ to 0.1 condenses the bound state to a single metastable state. The two learned metastable states at $\gamma$=0.1 also coincide with the bound and unbound states, with the border between the two marking the de-wetting transition. As with methane unbinding, discovering this metastable representation and entropy barrier corresponding to the de-wetting transition is only possible by directly accounting for the free-energy profile of the RC in the SPIB loss function.

Although we have discovered RCs with maximal entropy barriers to unbinding, we have not quantified how much each input feature contributes to the dissociation mechanism. Since the SPIB RCs are learned using a non-linear encoder, they are not readily interpretable. To interpret the important input features to the SPIB for transitioning between the metastable states identified using SPIB, we utilize the Thermodynamically Explainable Representations of AI and other black-box Paradigms (TERP) method\cite{Mehdi2023, Wang2023b}, which, given local inputs and their mapped RC values when passed through the non-linear encoder, finds the best local, linear approximation to the model by minimization of a loss function that contains an accuracy term regularized by a complexity loss similar in spirit to a Bayesian information criteria\cite{Bishop2006}; full details of the method can be found in Ref. \onlinecite{Mehdi2023}.

To determine which input features are the most important to the SPIB RC for describing transitions between the SPIB metastable states, we select 100 samples from the border between each metastable state. These are samples likely to belong to the transition state ensemble, as the boundary between SPIB states corresponds to isocomittor equaling 0.5\cite{Wang2021a,Beyerle2022}. In other words, these samples are points in the trajectory that transition to the neighboring metastable state in the subsequent timestep. Analyzing the local, linear model generated by TERP for these samples allows us to interpret the SPIB RC's behavior in that region, meaning TERP can be used to estimate which input features are important for describing the metastable transitions.

\begin{figure}[tp] 
\center
\includegraphics[width=0.9\linewidth]{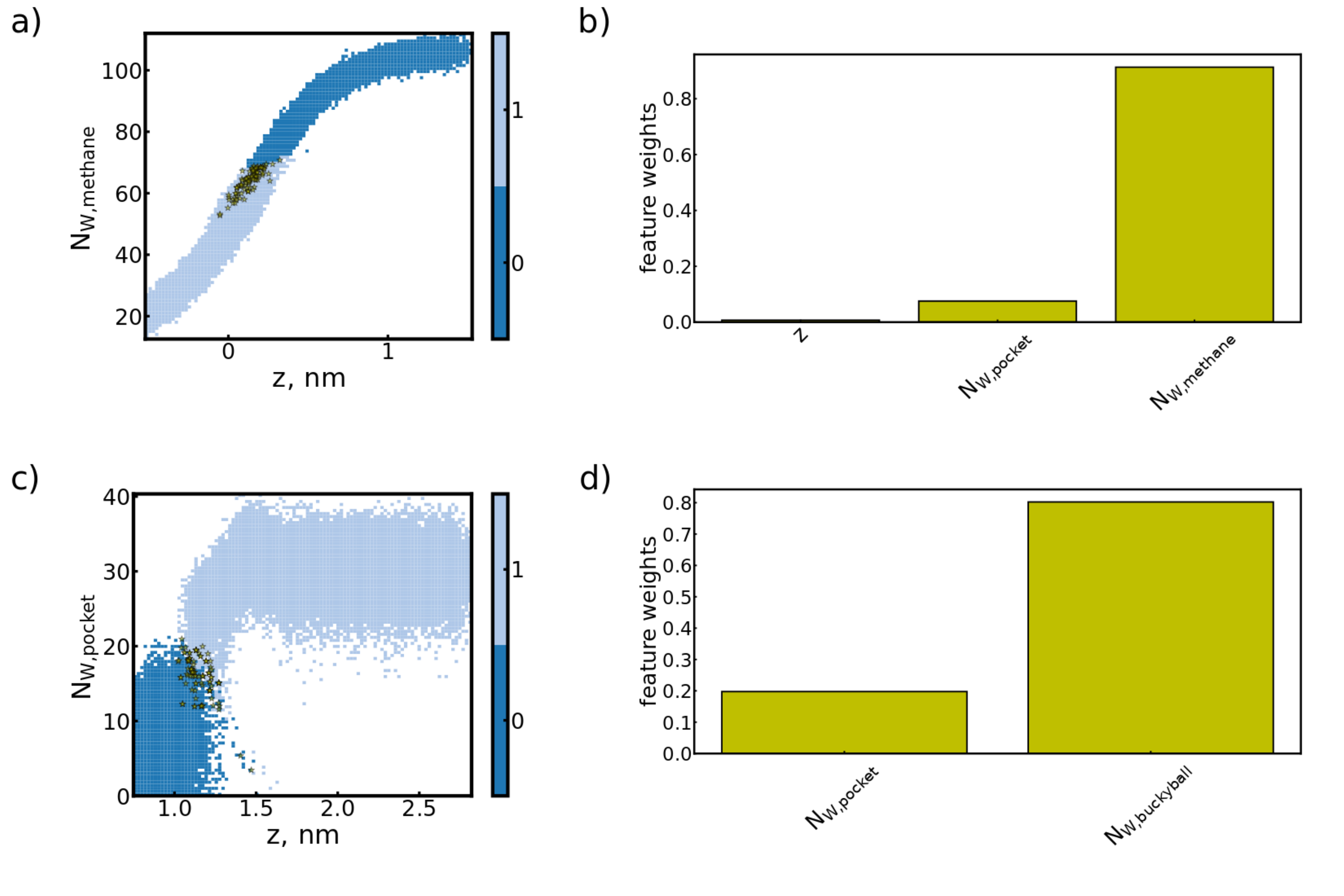}
\caption{a) Metastable states of the SPIB-learned RC for methane when $\gamma=0.1$, the same case as presented in Figure \ref{fig:spib_methane}, when projected onto the space spanned by a) (z, N$_{\text{W, methane}}$). The yellow stars denote 100 random samples from the border between the metastable states. These points are input to TERP to extract the relevant SPIB features for the transition between the metastable states. The non-zero feature weights for the best linear, local surrogate SPIB model are given in b). Error bars are negligible compared to bar height and have been omitted. c) Metastable states of the SPIB-learned RC for the fullerene system when $\gamma=0.1$, the same case as presented in Figure \ref{fig:spib_fullerene}, when projected onto the space spanned by (z, N$_{\text{W, pocket}}$). The yellow stars have the same function as in a). These points are input to TERP to extract the relevant SPIB features for the transition between the metastable states. The non-zero feature weights for the best linear, local surrogate SPIB model are given in d) The error bars in the bar plots indicate two standard errors of the mean calculated over the 100 sampled trajectory points.}
\label{fig:terp}
\end{figure}

For interpretation of the methane unbinding RC, we examine the SPIB model when the $\gamma$ hyperparameter is set to 0.1, corresponding to the results in Figure \ref{fig:spib_methane}. In this case, the SPIB model predicts there are two metastable states, shown in Figures \ref{fig:terp}a and \ref{fig:terp}b, corresponding to the bound and unbound states. The SPIB RC is interpreted at 100 samples on the border between the metastable states. The feature coefficients in the linear, local model for these points are shown in Figure \ref{fig:terp}b. Using this result, we see that transitions between the SPIB metastable states is dictated primarily by solvation of the methane particle, with solvation of the pocket playing a secondary role. Despite the z-distance between ligand pocket being the trivial reporter on ligand binding, at the transition state, z plays a nearly vanishing role. This effect is likely related to the sharp de-wetting transition \cite{Tiwary2016} seen previously for the fullerene equivalent of this system. There the transition to the bulk occurs due to wetting of the pocket, not a large-scale movement along z. Here, the transition states in the metastable region are methane poised at the mouth of the pocket (Figure S4), where small changes in the RC will lead to rapid solvation or desolvation of the methane. Thus, although changes in z are by necessity required for unbinding, they do not play a major role in the transition between bound and unbound states for this methane-pocket system.

The analogous TERP analysis is performed for the SPIB RC learned for the fullerene system with the $\gamma$ hyperparameter set equal to 0.1; the results are displayed in Figure \ref{fig:terp}c,d. For the fullerene system the SPIB discovers two metastable states, Figure \ref{fig:terp}c, with the weight of the input features at the transition state given in Figure \ref{fig:terp}d. Similar to the methane system, the main driver of transitions between bound and unbound states is the solvation of the hydrophobic ligand, Interestingly, despite the more significant change in the pocket hydration state when the fullere is unbound, the solvation of the pocket plays a secondary role in SPIB RC. This result of the pocket solvation playing the minor role in the RC is in agreement with previously learned linear RCs for this system \cite{Tiwary2016wet,Ribeiro2018}. However, in contrast to these previous studies, we explicitly account for the solvation state of the solute in the learned RC.

While it may seem that the lack of weight of the z coordinate is in conflict with the RCs reported in Refs. \onlinecite{Tiwary2016wet,Ribeiro2018}, we note that we only examine the importance of the input features to the RC \emph{belonging to the transition state ensemble} whereas the RCs elucidated in Refs. \onlinecite{Tiwary2016wet,Ribeiro2018} are reporting on the \emph{global importance} of the input features to the RC by examining the weights of a \emph{linear} model. The global importance of the z-coordinate to the SPIB RC for the fullerene system can be seen by examination of the projection of the RC in the (z, N$_{\text{W, pocket}}$) space in Figure \ref{fig:spib_fullerene}a, where the projected RC increases as a function of z.

Here we have reported machine-learned reaction coordinates (RCs) describing the dissociation of two model hydrophobic ligands, a united atom methane particle and a C$_{60}$ fullerene, of different sizes from hydrophobic pockets with slightly different geometries but identical non-bonded interaction potentials. We take the established state predictive information bottleneck (SPIB) method \cite{Wang2021a} for learning RCs and add an extra penalty term to the loss function to encourage the discovery of RCs possessing free energy profiles with large entropy barriers. While the entropy profiles along the learned RCs for both systems are qualitatively similar, the energy profiles differ significantly.

For unbinding of the united atom methane, the optimized RC shows that the methane must first surmount a small energy barrier followed by a larger entropy barrier. The unbinding process is uphill along the RC's free energy profile until the solute reaches the bulk solvent. For the optimized coordinate, we find two metastable states, one for the bound state and one for the unbound state. The border between them is demarcated by a large entropy barrier. Adding the extra penalty term to the loss function explicitly accounting the for the entropy barrier along the RC is required for discovering this optimized RC with a non-trivial unbinding thermodynamic profile. 

For the C$_{60}$ fullerene, the optimized RC indicates that unbinding from the pocket is energetically favorable, but is overall uphill in free energy along the RC, due to a large entropy cost. As with methane, the unbinding free-energy barrier is entropy dominated. The unbinding free energy change is significantly larger for the fullerene compared to the methane particle, roughly 75 kJ / mol for the fullerene versus roughly 15 kJ / mol for methane. 

It should be emphasized that the results presented here for both methane and the fullerene systems rely on calculating the energy profile via the short-ranged intermolecular interactions of the ligand with the rest of the system. As such, the entropy barriers to unbinding in both cases are likely due to the ligand being able to form a larger number of more favorable intermolecular interactions with the TIP4P water solvent, which has a larger Lennard-Jones well-depth parameter compared to the hydrophobic pocket atoms. If the type or the parameterization of the solvent is changed, the discovered RC and subsequent thermodynamic profiles will change as well.

These results demonstrate the utility of including an explicit thermodynamic penalty when using machine learning for the discovery of reaction coordinates for non-trivial physical systems. Without the extra term describing the entropy barrier along the RC in eq. \ref{eq:loss2}, the SPIB approach is unable to learn an RC possessing a significant entropy barrier to dissociate for either system. This type of entropy-dominated reaction coordinate should be useful as a biasing variable for use in path-based enhanced sampling approaches such as milestoning \cite{Faradjian2004}, transition path sampling\cite{Swenson2019}, and forward flux sampling \cite{Allen2005}. We also expect the learned RCs presented here to be transferrable to more realistic models of hydrophobic binding and unbinding, such as the noted interactions of C$_{60}$ fullerenes with certain proteins \cite{Calvaresi2010,Calvaresi2014,Calvaresi2015} or drug dissociation from proteins \cite{Bonati2021} and RNA\cite{Levintov2020}.

\begin{acknowledgement}

The authors thank the members of the Tiwary Lab for useful discussions over the course of this research project, especially Shams Mehdi for his assistance with the TERP analysis. This work is entirely funded by the US Department of Energy, Office of Science, Basic Energy Sciences, CPIMS Program, under Award DE-SC0021009.

\end{acknowledgement}

\begin{suppinfo}

More information regarding the parameters for the well-tempered metadynamics simulations and additional SPIB analysis on the methane and fullerene systems can be found in the supporting information, available free of charge at XXX.

\end{suppinfo}

\bibliography{achemso}

\end{document}


\section{Molecular Dynamics Simulation Details}
\begin{figure}[tp] 
\center
\includegraphics[width=0.9\linewidth]{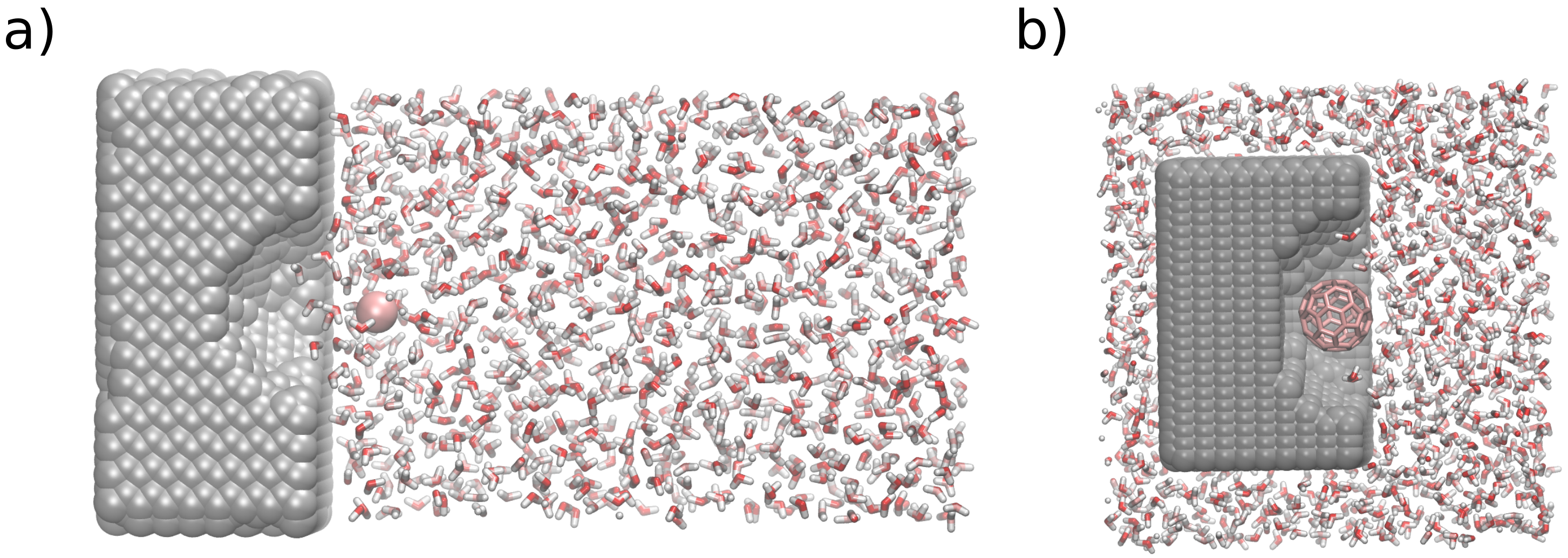}
\caption{Transverse cuts along the x-axis of the starting structures for the a) methane-pocket system and b) the fullerene-pocket system. In both cases, the particles composing the hydrophobic pocket are represented as grey spheres, the ligand is represented as either a pink sphere (methane) or in the licorice representation (fullerene). All TIP4P water atoms are shown in licorice representation with oxygens in red and hydrogens in white and the virtual sites omitted. The relative size of the systems to each other is not to scale. Both images are created using VMD v1.9.3\cite{Humphrey1996}.}
\label{fig:snapshots}
\end{figure}

Transverse cuts of the starting structures of the methane and fullerene systems are given in Figure \ref{fig:snapshots}.
\subsection{Methane}
The initial structure for this system is generated by first creating a 35.0 $\text{\AA}$ $\times$ 33.0 $\text{\AA}$ $\times$ 100.0 $\text{\AA}$ simulation box in LAMMPS, with a slab of paraffin-like particles of volume 35.0 $\text{\AA}$ $\times$ 33.0 $\text{\AA}$ $\times$ 15.0 $\text{\AA}$ in an HCP lattice created at one end of the box. To create the pocket in the slab, all particles within a sphere of radius 8.0 $\text{\AA}$ centered at 17.5 $\text{\AA}$ $\times$ 16.5 $\text{\AA}$ $\times$ 15.0 $\text{\AA}$ are removed using the $\mathsf{delete}$\_$\mathsf{atoms}$ command in LAMMPS. These paraffin-like particles interact with the methane ligand and the solvent only through the Lennard-Jones interaction parameterized using the values given in Table 1 in the main text. The locations of the particles in the slab are frozen using the GROMACS $\mathsf{freezegrps}$ option and periodic boundary conditions are applied in the x-,y-, and z-directions of the simulation box. However, since the hydrophobic slab is frozen translationally and occludes completely one end of the simulation box, there is no mass transport across the z-direction. TIP4P waters are added using the GROMACS $\mathsf{insert}$-$\mathsf{molecules}$ command, which results in the addition of 1931 solvent molecules to the box, corresponding to a density of $\sim$979 kg/m$^3$ in the bulk. The system is energy minimized using the steepest descent algorithm, then subject to a 2-ns equilibration in the NVT ensemble using the velocity-rescaling thermostat \cite{Bussi2007} set to 300 K; no position restraints were utilized. To prevent the ligand from drifting too far into the bulk water, a quartic restraining position is placed at $z$= 1.1 nm from the mouth of the pocket during the production run described in the main text. The production run analyzed in the main text is a total of 76.1 ns in length in unbiased simulation time.

\subsection{Fullerene}
The parameterization for this system is the same as used previously\cite{Tiwary2015,Ccoa2022,Ahalawat2020}. The hydrophobic pocket and fullerene are solvated with 3375 TIP4P water molecules, and the density of the water in the bulk of the simulation box is 1004 kg/m$^3$. A single simulation of this systems is run for $\sim$ 94.2 ns of unbiaed simulation times, which is significantly longer than simulations previously used to converge the free energy \cite{Ccoa2022} and reaction coordinate \cite{Tiwary2016wet,Ribeiro2018}. WTMetaD biasing parameters are identical to those used in \cite{Ccoa2022} for the biased simulations with no applied force, although we have used a stronger force constant for the quartic wall potential that prevents the fullerene from diffusing too far from the pocket. 
\subsection{Well-tempered Metadynamics parameters}
Table \ref{tab:WTMetaD} gives the well-tempered metadynamics parameters for the biased simulations the methane and buckyball analyzed in the main text.
\begin{table}[b]
    \centering
    \caption{Parameters for WTmetaD}
    \begin{tabular}{cccccc}
    
        System & $\omega$(kJ/mol) & $\gamma$  & $\sigma_z$, nm & $\tau$, ps &$T$ (K) \\ 
        \hline
        
         Methane & 2.0 & 15 & 0.03 & 0.6 & 300  \\

         Fullerene & 2.0 & 15 & 0.03 & 0.6 & 300 \\

    \end{tabular}
    \label{tab:WTMetaD}
\end{table}
\section{Pocket Solvation as a Function of Ligand Distance}
\begin{figure}[tp] 
\center
\includegraphics[width=0.9\linewidth]{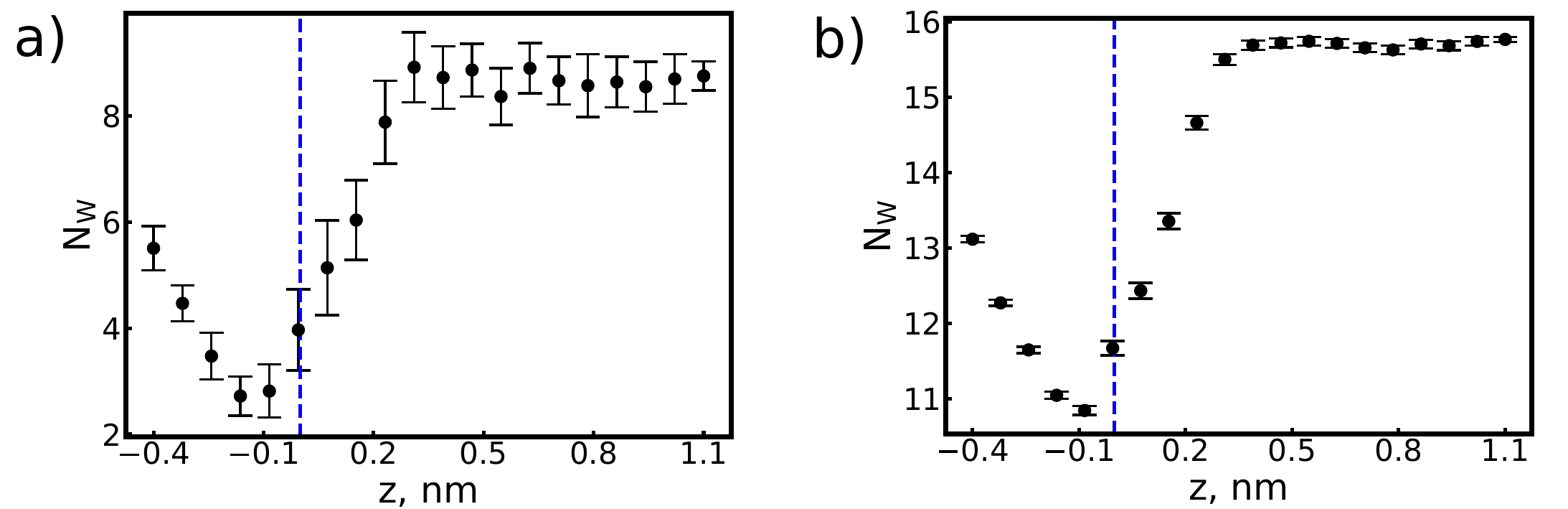}
\caption{Number of waters in the pocket as a function of ligand z-distance from the plane of the pocket lip for the methane simulation, determined by a) counting the number of water oxygen atoms with z $\le$ 0.0 nm and b) using the coordination number, as implemented in PLUMED 2, of an atom at the bottom of the pocket with a cutoff radius of 0.8 nm. Dashed, blue line indicates the mouth of the pocket.}
\label{fig:CN_methane}
\end{figure}

Figure \ref{fig:CN_methane} shows the water population of the hydrophobic pocket in the methane simulation calculated using the two different approaches. In Figure \ref{fig:CN_methane}a, the number of waters in the pocket is calculated using the method described in Ref. \onlinecite{Baron2012}, where all the oxygen atoms of waters at z values less than 0.0 nm were counted as inside the pocket while in Figure \ref{fig:CN_methane}b the counting is performed using the coordination number collective variable, $\mathsf{COORDINATION}$, in PLUMED with a cutoff radius of 0.8 nm. Qualitatively the two curves are identical, differing only quantitatively in the number of waters present in the pocket at each value of z. This qualitative agreement indicates that using the coordination number gives the correct wetting behavior of the pocket as the ligand unbinds. 

The pocket is minimally hydrated when the methane ligand is approximately 0.1 nm inside the pocket, which is likely due to the consistent formation of a hydrophobic vacuum between the bottom of the pocket and the methane. However, the wetting of the pocket is non-monotonic, as, once the methane reaches the bottom of the pocket, the pocket re-hydrates, although not to the extent of methane in the bulk solvent. In agreement with previous work, the rehydration process of the pocket following methane unbinding is monotonic.

\begin{figure}[tp] 
\center
\includegraphics[width=0.9\linewidth]{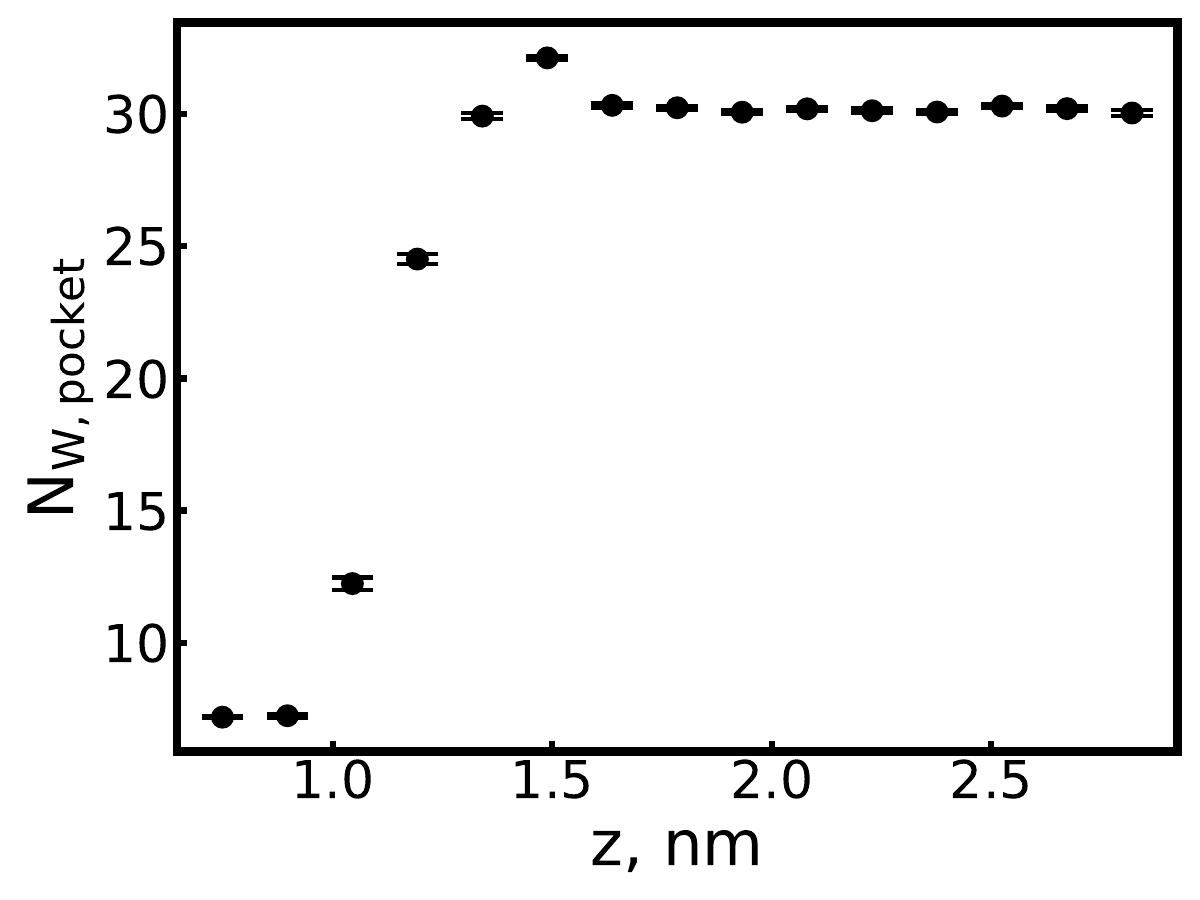}
\caption{Number of waters in the pocket as a function of ligand z-distance from the plane of the pocket lip for the buckyball simulation, determined by using the coordination number, as implemented in PLUMED 2, of an atom at the bottom of the pocket with a cutoff radius of 0.8 nm.}
\label{fig:CN_buckyball}
\end{figure}

The behavior of the pocket hydration upon unbinding in the case of the C60 fullerene is quantitatively different. Figure \ref{fig:CN_buckyball} shows that hydration of the pocket is monotonic with solute distance from the pocket until the fullerene reaches the bulk state, where there is a slight maximum near z = 1.5 nm. That is, the buckyball is large enough and the shape of the pocket is such that, as the fullerene unbinds, waters continuous fill the pocket. With the smaller methane ligand, the pocket must dehydrate as the methane moves toward the middle of the pocket, then rehydrate as it exits, due to the smaller size of the methane, which allows more waters to partially fill the pocket as the methane reaches the pocket's bottom.

\section{Snapshots from the Entropy Barrier along the SPIB RC for Methane and Fullerne}
\begin{figure}[tp] 
\center
\includegraphics[width=0.9\linewidth]{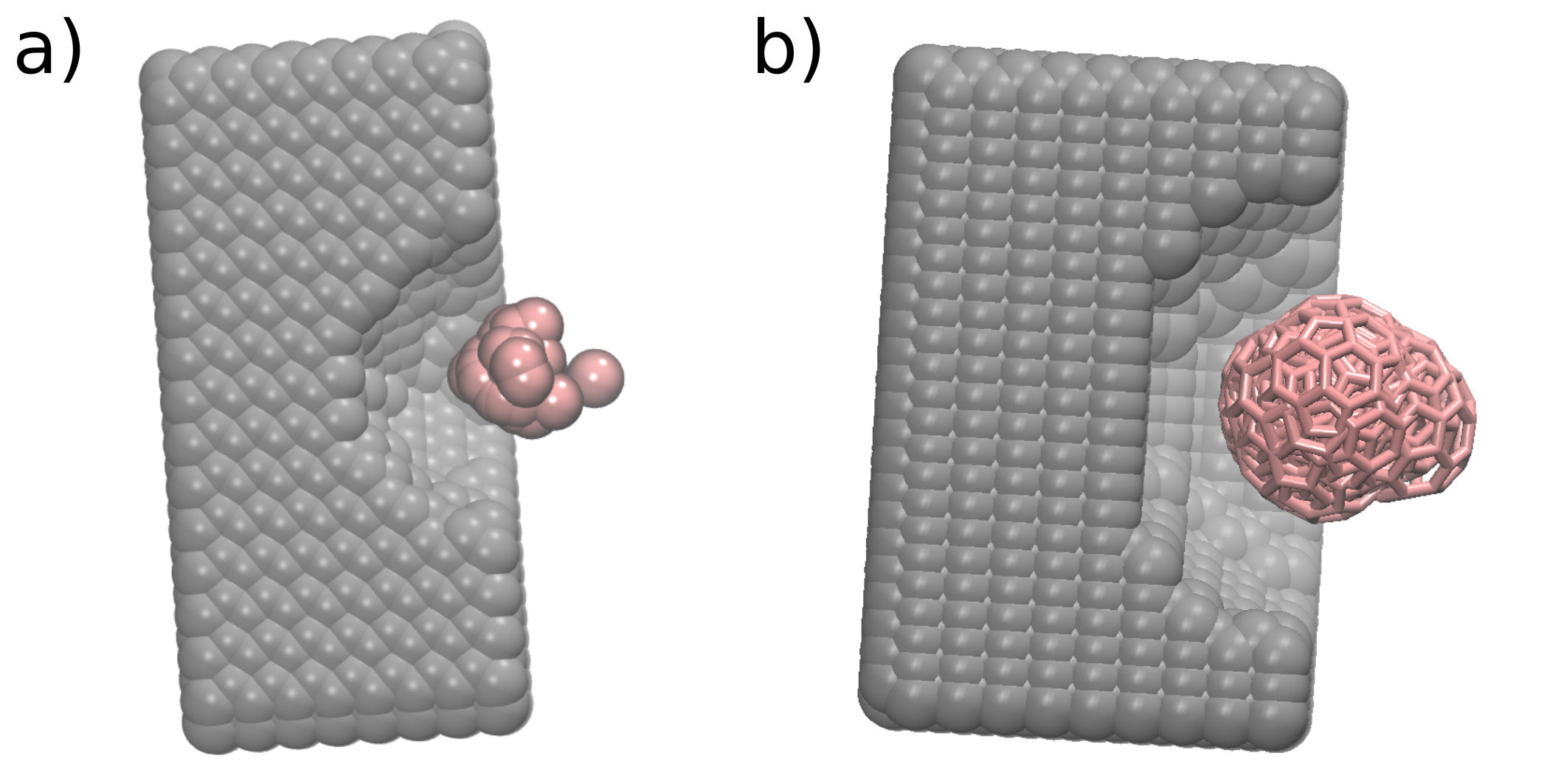}
\caption{Snapshots of a) methane and b) fullerene when the respectively simulation trajectory samples the neighborhood of the peak of the entropy barrier.}
\label{fig:snapshots_barrier}
\end{figure}

Figure \ref{fig:snapshots_barrier} shows the position of the ligand for around 100 frames sampled from the peak of the entropy barrier for methane, Figure \ref{fig:snapshots_barrier}a, and the fullerene, Figure \ref{fig:snapshots_barrier}b. In each case, crossing the entropy seems to occur when the ligand is poised on the threshold of the pocket.
\section{SPIB Metastable States as a Function of the $\gamma$ Hyperparameter}
\begin{figure}[tp] 
\center
\includegraphics[width=0.9\linewidth]{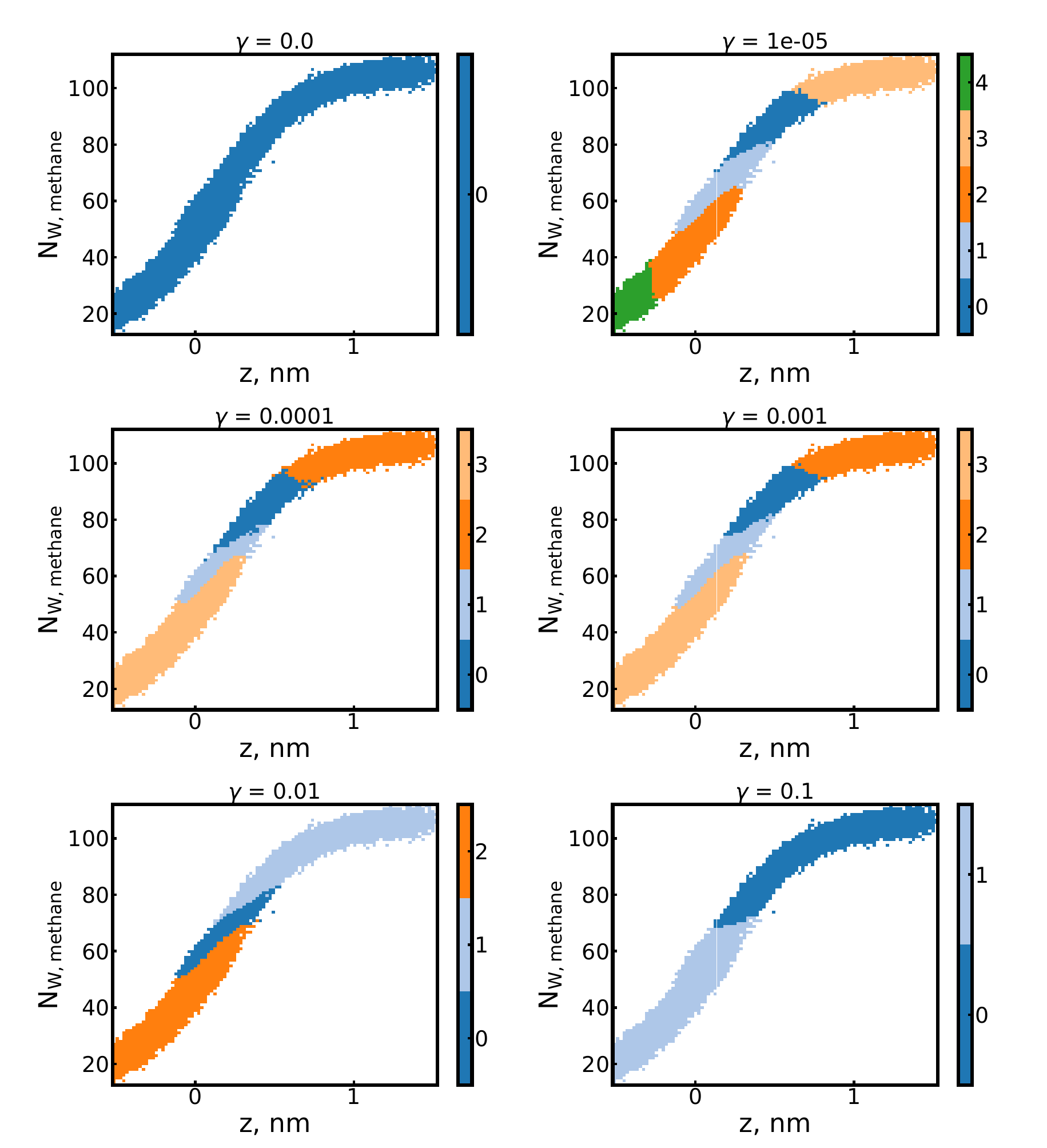}
\caption{Number and topology of the SPIB-learned metastable states for the methane unbinding system as a function of $\gamma$, indicated above each of the respective subplots. All other SPIB hyperparameters are the same for each result.}
\label{fig:metastable_states}
\end{figure}

\begin{figure}[tp] 
\center
\includegraphics[width=0.9\linewidth]{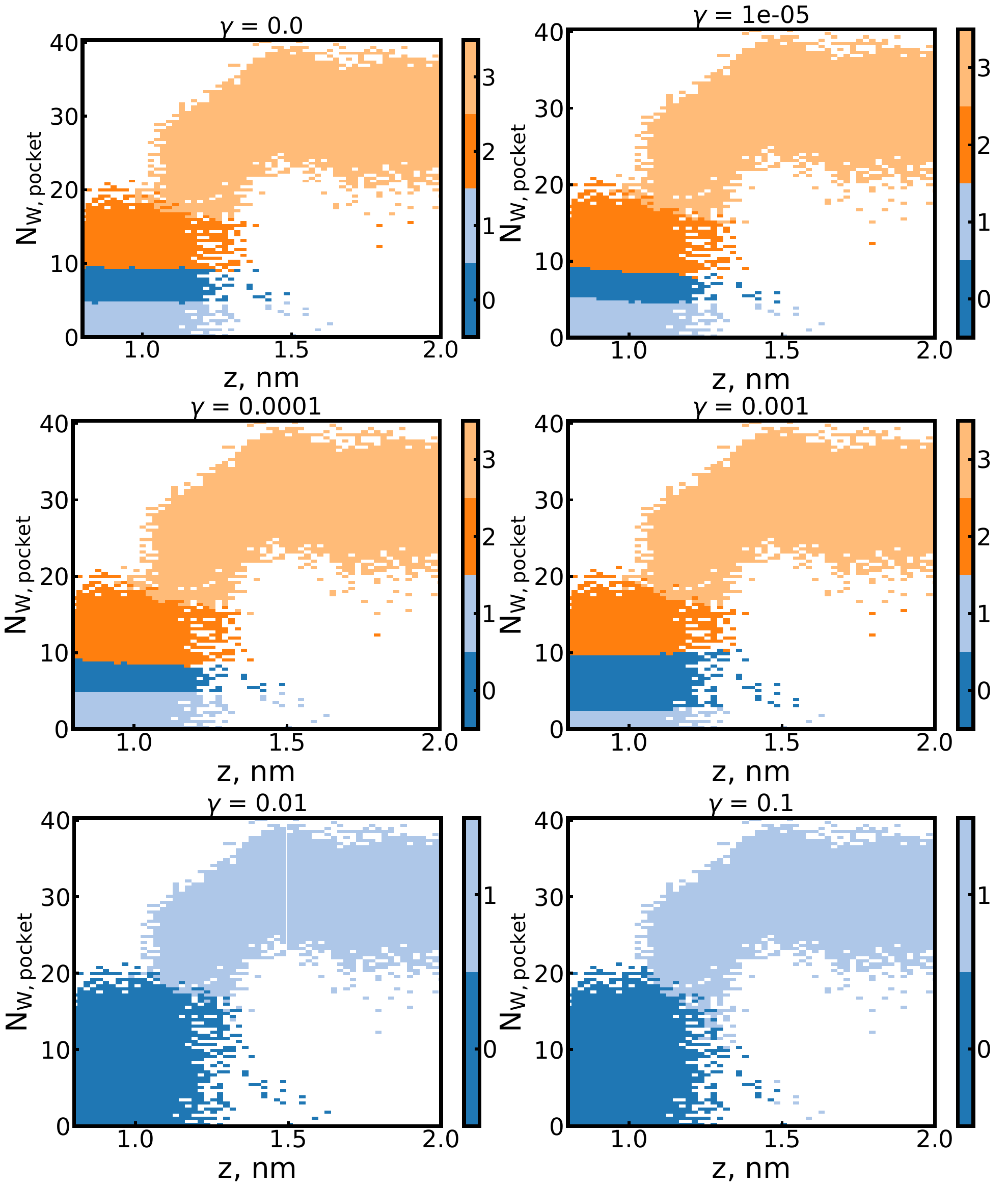}
\caption{Number and topology of the SPIB-learned metastable states for the buckyball unbinding system as a function of $\gamma$, indicated above each of the respective subplots. All other SPIB hyperparameters are the same for each result.}
\label{fig:fullerene_metastable}
\end{figure}

Figure \ref{fig:metastable_states} shows the number and shape of the metastable states predicted by SPIB as a function of the $\gamma$ of the loss augmented SPIB loss function given in eq. 2 of the main text. When there is no extra penalty, $\gamma$ = 0.0, the SPIB predicts a single metastable state due to the lack of a sufficient barrier along the learned RC. Addition of even a small penalty, $\gamma=1e-5$, allows for a much more detailed decomposition of the relevant (z, N$_{\text{W, methane}}$) input space. Further increasing $\gamma$ allows for collapse of the learned metastable states into two states, the bound and unbound states, due to the increasing sharpness of the learned entropy barrier along the reaction coordinate. For all model results shown in Figure \ref{fig:metastable_states} all other SPIB hyperparameters are kept constant excepting $\gamma$. The analogous plots are given for the fullerene system in Figure \ref{fig:fullerene_metastable}; in contrast to the methane system, there are a non-trivial number of metastable states even for the case when $\gamma$ = 0.0.

\newpage
\bibliography{achemso}